\documentclass[aps, twocolumn, nofootinbib, showpacs, prd]{revtex4}
\usepackage{graphicx}
\usepackage{amsfonts}
\usepackage{amssymb}
\usepackage{amsbsy}
\usepackage{amsmath}
\usepackage{mathtools}
\usepackage{latexsym}
\usepackage{natbib}
\usepackage{bm}
\usepackage{ulem}
\usepackage[utf8]{inputenc}
\usepackage{newunicodechar}
\newunicodechar{∼}{$\sim$}
\usepackage{color}
\usepackage{float}
\usepackage{hyperref}
\usepackage{multirow}
\usepackage{array,makecell}

\begin{document}
\title{Gravitational Collapse of a Chiellini Integrable Scalar Field}

\author{Mohamed Aarif A}
\email{mohamedaarif.a2023@vitstudent.ac.in}
\affiliation{School of Advanced Sciences, Vellore Institute of Technology, \\ 
Tiruvalam Rd, Katpadi, Vellore, Tamil Nadu 632014 \\ India}

\author{Soumya Chakrabarti}
\email{soumya.chakrabarti@vit.ac.in}
\affiliation{School of Advanced Sciences, Vellore Institute of Technology, \\ 
Tiruvalam Rd, Katpadi, Vellore, Tamil Nadu 632014 \\ India}

\pacs{}

\date{\today}

\begin{abstract}
We study the gravitational collapse of a non-interacting mix of perfect fluid and a spatially homogeneous scalar field within a Chiellini-integrable framework. We choose an extended Higgs-type self-interaction potential and reduce the Klein-Gordon equation into a generalized damped Milne-Pinney class of differential equation. We derive a closed-form analytical solution for the scalar field, the scale factor and explore the collapsing branch of the same. We find that it exhibits an asymptotic collapse in which the proper volume decreases monotonically but never reaches zero at finite time. We analyze the energy conditions for the constituent elements of the collapsing sphere. While the scalar field remains canonical in nature, we find that the perfect fluid can violated the Null Energy Condition. We also study the formation of apparent horizon condition and find multiple possibilities depending on the parameter space : either no trapped surface or the formation of multiple apparent horizons. We match the interior homogeneous solution to a generalized Vaidya exterior via the Israel-Darmois junction conditions, yielding the corresponding boundary mass function, ensuring a smooth collapse scenario.
\end{abstract}

\maketitle

\section{Introduction}
\label{sec:introduction}
Non-linear differential equations are fundamental to modern mathematical physics. Many physically relevant systems exhibit a degree of non-linearity that cannot be captured within linear approximations. This is particularly evident in General Relativity (GR) and its extensions, where the field equations connect spacetime geometry with matter fields and allows one to explore gravitational interactions. In the standard theory as well as its modifications, the non-linearity of the field equations severely limits the availability of exact analytical solutions, especially in dynamical settings. A popular research problem where this non-linearity becomes an integral part of the system is gravitational collapse, usually associated to the fate of a massive stellar distribution after it has exhausted all source of internal pressure. The collapse problem remains one of the most fundamental problems of gravitational physics yet to be fully resolved. It is expected that a self-gravitating matter distribution should collapse under its own gravity if there is no source of internal pressure and drive the system towards a strong curvature limit and culminate in the formation of a spacetime singularity. The simplest and the very first model of this phenomenon, namely, the Oppenheimer-Snyder collape describes a spherically symmetric pressure-less dust and indeed provides an analytical solution \cite{datt, os}. Standard models of collapse often come with a formation of singularity and apparent horizon, which signals that the singularity is covered, an onset of black hole formation \cite{penrose1}. The ultimate fate of such a process is also tied to issues such as the cosmic censorship \cite{penrose2}. For a more realistic collapse model one must explore more generalized energy-momentum tensor and the resulting mathematical setup naturally becomes dependent on a set of more non-linear and coupled differential equations. \\

A widely studied class of collapse models involves scalar fields minimally or non-minimally coupled to gravity \cite{goncalves}. In the case of a \textit{massless scalar field}, the collapse can exhibit critical phenomena, where fine-tuning of initial conditions leads to universal scaling laws and self-similarity right before the formation of a black hole \cite{choptuik, brady, gundlach}. For a \textit{massive scalar field}, the presence of a self-interaction potential leads to modified field equations and as a consequence, modified solutions, exhibiting oscillatory collapse, dispersion/bounce or dynamical-equilibrium/bound states \cite{christ, piran}. Generalized models of scalar field collapse involves consideration of non-canonical kinetic terms or non-minimal coupling with regular matter which naturally extends the scope of the system. There has been considerable development in these approaches, particularly through numerical relativity, as well as using symmetry-reduced systems. However, in general the systems remain non-trivial due to their non-linearity and an exact treatment is diffucult to come across. This serves as a motivation to look for mathematical frameworks where one can tackle the non-linear features of the system analytically. \\

We identify that the Milne-Ermakov-Pinney (MEP) system of equations can provide such a framework. The equation was originally explored by Ermakov \cite{Ermakov} and later extended by Milne and Pinney \cite{Milne, Pinney}. They introduced a non-linear auxiliary equation in order to construct invariants associated with the time-dependent harmonic oscillator. This led to the construction of an \textit{Ermakov-Lewis invariant}, using which one can identify the non-linear system of equations that can admit exact solutions. The MEP equation has already been used in a wide range of physics problems, such as quantum mechanics, non-linear optics, plasma dynamics and kinematic evolution of fluids \cite{Haas, Carinena}. The standard form of the equation can be written as
\begin{equation}
\ddot{y} + \omega^2(t) y = \frac{\kappa}{y^3},
\label{eq:MEP}
\end{equation}
which can easily be interpreted as an extension of the linear oscillator equation $\ddot{x} + \omega^2(t)x = 0$. $\omega(t)$ works as a time-dependent frequency and $\kappa$ is a real parameter. Eq.~\eqref{eq:MEP} describes an idealized system; a more realistic physics problem often comes with differential equations with dissipative terms. Introducing a linear damping term to the above equation leads to the \textit{damped Milne-Ermakov-Pinney equation}, written as
\begin{equation}
\ddot{y} + \mu \dot{y} + \omega^2(t)y = \frac{\kappa}{y^3}, \qquad \mu > 0,
\label{eq:DMEP}
\end{equation}
where $\mu$ denotes the damping coefficient. It is known that through suitable transformations, Eq.~\eqref{eq:DMEP} can be mapped into a generalized Emden-Fowler equation, which allows an extraction of exact solution. Rogers \cite{Rogers} showed that a hybrid Ermakov-Painleve system may arise from a reduction of a coupled nonlinear Schrodinger equation system. A similar reduction can be found in the theory of solitons and non-linear wave equation. In extension to these works, there is a renewed interest in Chiellini-type integrability conditions \cite{AC}. These conditions establish a specific functional relation between the damping and restoring terms in a Lienard-type differential equation and guarantees exact integrability \cite{AL}, even in dissipative systems \cite{IB1, IB2, MH1, MH2, MH3, YY, MH4, RMC, MR, MR1}.  \\

We explore the role of Chiellini integrability condition in GR, in particular in a spherically symmetric collapse of a scalar field. The damped Ermakov-Milne-Pinney equation and the resulting integrability has been explored in the past, in the context of a self-similar scalar field collapse \cite{nbsc}, however, the scope of a Chiellini integrability remains unexplored. We show that the Klein-Gordon equation for a spatially homogeneous scalar field minimally coupled to GR can be easily reformulated as a damped Ermakov-Painleve-II equation. The resulting system admits closed-form analytical solutions through a Chiellini-integrable framework. We demonstrate that these solutions describe a homogeneous gravitational collapse in which the areal radius never reaches a zero proper-volume at any finite time. We explore the associated energy conditions of the collapsing fluid element. We derive the condition for the formation of apparent horizon and find out that depending on the parameter space, there is either no formation of horizon or a formation of multiple horizons.

\section{Framework : A Chiellini-Integrable Scalar Field Collapse in General Relativity}
First, we summarize the integrability properties of the \textit{damped Ermakov--Painlev\'e~II} and the \textit{generalized damped Milne-Pinney (GDMP)} equations under the Chiellini integrability condition. These systems provide a concrete realization of nonlinear dynamics in which dissipative and restoring effects coexist while preserving exact analytic solvability.

\medskip
\noindent\textbf{Integrability.}
\textit{
Consider the damped Ermakov--Painlev\'e~II equation
\begin{equation}
    \ddot{y} + g(y)\dot{y} + h(y) = 0, 
    \qquad h(y) = \lambda y + \epsilon y^3 - \frac{\eta}{y^3},
    \label{eq:DEP}
\end{equation}
If the Chiellini condition
\[
    \frac{d}{dy}\!\left(\frac{h(y)}{g(y)}\right) = R_{0}\, g(y), 
    \qquad R_{0} \in \mathbb{R},
\]
is satisfied, the equation becomes analytically solvable. Representative exact solutions are
\[
    y(t) = \sqrt{\frac{1}{(t-t_0)^2} + \sqrt{\tfrac{p}{3}}}, 
    \qquad \epsilon = -1,
\]
and
\[
    y(t) = 
    \left(\frac{p^2}{16\lambda^2} - \frac{\eta}{\lambda}\right)^{\!1/2}
    \sin\!\big[2\sqrt{2\lambda}(t-t_0)\big] 
    + \frac{p}{4\lambda}, 
    \qquad \epsilon = 0.
\]
}

Equation~\eqref{eq:DEP} incorporates three distinct nonlinear contributions: a cubic Painlev\'e-type term $\epsilon y^3$, a linear restoring term $\lambda y$, and the inverse-cubic Ermakov term $-\eta/y^3$. The damping function $g(y)$ is not arbitrary; the Chiellini condition imposes a specific functional relation between $g(y)$ and $h(y)$, effectively reducing the second-order equation to a separable first-order form. This structural constraint guarantees integrability even in the presence of nonlinear dissipation.  \\

For $\epsilon=-1$, the dynamics exhibit an inverse-square-root decay, whereas $\epsilon=0$ yields oscillatory behaviour with effective frequency $\sqrt{2\lambda}$. The inverse-cubic term modifies both amplitude and phase, leading to parameter-dependent nonlinear frequency shifts. These regimes illustrate how a unified dynamical structure can generate qualitatively distinct physical evolutions.  \\

We consider a scalar field minimally coupled to the Einstein-Hilbert action
\begin{equation}\label{action1}
\textit{A}=\int{\sqrt{-g} d^4x [R + \frac{1}{2}\partial^{\mu}\phi\partial_{\nu}\phi - V(\phi) + L_{m}]}.
\end{equation}
Here $L_{m}$ denotes the Lagrangian density of ordinary matter. Variation with respect to the metric yields the scalar-field energy-momentum tensor
\begin{equation}\label{minimallyscalar}
T^\phi_{\mu\nu} = \partial_\mu\phi\partial_\nu\phi - g_{\mu\nu}\Bigg[\frac{1}{2}g^{\alpha\beta}\partial_\alpha\phi\partial_\beta\phi - V(\phi)\Bigg]. 
\end{equation}

For a spatially homogeneous scalar field in a homogeneous and isotropic spatially flat spacetime (with $8\pi G = 1$), described by
\begin{equation}
ds^2 = -dt^2 + a^2(t)\left[dr^2 + r^2 d\Omega^2 \right],
\end{equation}
the field equations reduce to
\begin{equation} \label{fe1minimal}
3\Big(\frac{\dot{a}}{a}\Big)^{2} = \rho_{m} + \rho_{\phi} = \rho_{m} + \frac{\dot{\phi}^{2}}{2} + V\left( \phi \right),
\end{equation}
\begin{equation} \label{fe2minimal}
-2\frac{\ddot{a}}{a}-\Big(\frac{\dot{a}}{a}\Big)^{2} = p_{m} + p_{\phi} = p_{m} + \frac{\dot{\phi}^{2}}{2}-V\left( \phi \right),
\end{equation}
together with the Klein--Gordon equation
\begin{equation} \label{phiminimal}
\ddot{\phi} + 3\frac{\dot{a}}{a}\dot{\phi} + \frac{dV(\phi)}{d\phi} = 0.  
\end{equation}

For the self-interaction potential
\begin{equation}\label{ourpot}
V(\phi) = V_0 + \frac{\lambda}{2}\phi^2 + \frac{\epsilon}{4}\phi^4 + \frac{\eta}{2 \phi^2}, 
\end{equation}
the scalar evolution equation becomes
\begin{equation}\label{minKG}
\ddot{\phi} + 3\frac{\dot{a}}{a}\dot{\phi} + \lambda \phi + \epsilon \phi^3 - \frac{\eta}{\phi^3} = 0,
\end{equation}
which belongs to the Chiellini-integrable class. The quadratic and quartic terms in the self-interaction potential resemble Higgs-type scalar potentials popularly employed in field-theoretic contexts, such as running vacuum models and symmetrons \cite{jimenez, horacio, lopez, khoury, sola, cai, sckd}. The additional inverse-power contribution introduces a non-polynomial self-interaction reminiscent of effective potentials arising in axion-like or quantum-corrected scalar models \cite{ruffo, moore}. Together, these terms generate a nonlinear hierarchy that becomes particularly relevant in dynamical collapse scenarios. In the large-field regime $|\phi| \gg 1$, the quartic interaction dominates,
\begin{equation}
V(\phi) \simeq \frac{\epsilon}{4}\phi^4,
\end{equation}
and the scalar behaves as a strongly self-interacting component. The corresponding energy density and pressure,
\begin{equation}
\rho_\phi = \frac{\dot{\phi}^2}{2} + V(\phi),
\qquad
p_\phi = \frac{\dot{\phi}^2}{2} - V(\phi),
\end{equation}
imply that the effective equation-of-state parameter
\begin{equation}
w_\phi = \frac{\dot{\phi}^2 - 2V(\phi)}{\dot{\phi}^2 + 2V(\phi)}
\end{equation}
depends sensitively on the balance between kinetic and potential energies. Introducing $x=t-t_0$, the exact analytic solution can be written as
\begin{eqnarray}\label{eq:phi_of_t_general}
&& \phi(x) = \sqrt{\frac{1}{x^2} + \sqrt{\tfrac{p}{3}}}, \\&& \label{eq:a_of_t_general}
a(t) = c_1\, \exp\big\lbrace F(x)\big\rbrace \,\big(\sqrt{3} + \sqrt{p}\,x^2\big)^{E}\,x^{1/3}, \\&& 
E = \frac{1}{6} - \frac{\sqrt3\,\eta}{2\,p^{3/2}}, \\&&\nonumber
F(x) = -\frac{1}{108 p}\Big[-6(9\eta+3\lambda p-2\sqrt3\,p^{3/2}) \\
&& + \sqrt3\sqrt p(9\eta -3\lambda p +\sqrt3 p^{3/2})\,x^2\Big]x^2.
\end{eqnarray}

The parameters $(c_1, \eta, \lambda, p)$ determine the qualitative behaviour of the areal radius.  \\

\begin{figure}[t!]
\begin{center}
\includegraphics[angle=0, width=0.40\textwidth]{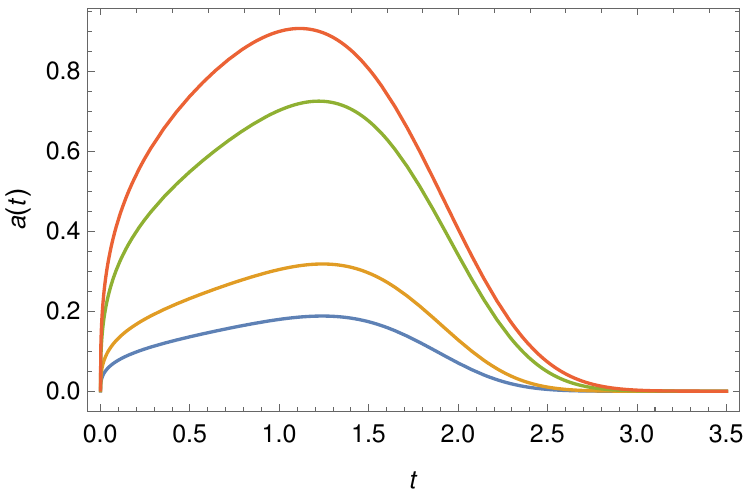}
\includegraphics[angle=0, width=0.40\textwidth]{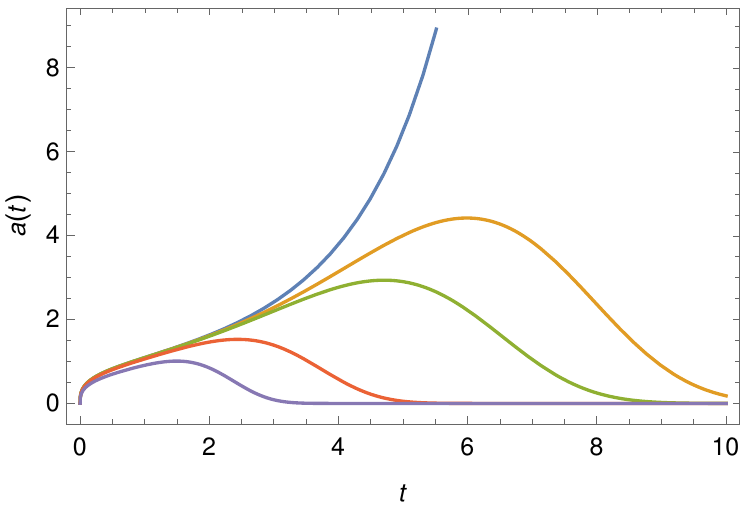}
\includegraphics[angle=0, width=0.40\textwidth]{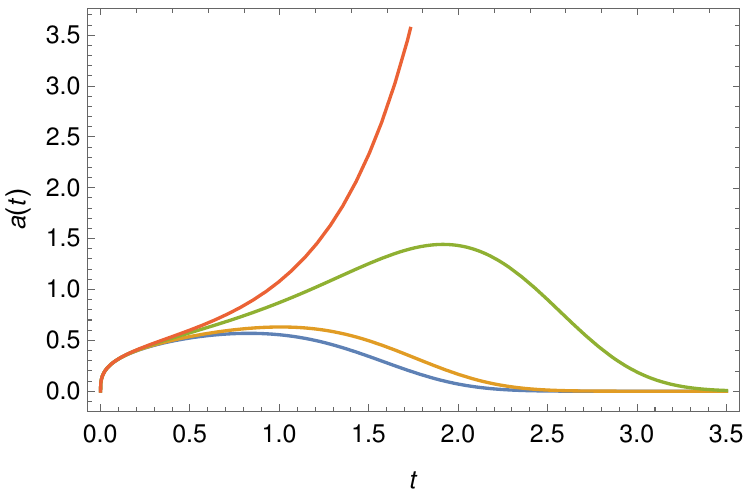}
\caption{$a(t)$ as a function of $t$ (taking $t_{0} = 0$). The graph on top is for different values of $p$, while $\eta$ and $\lambda$ are kept fixed. The graph in the middle is for different values of $\eta$, while $p$ and $\lambda$ are kept fixed. The bottom graph is for different values of $\lambda$}
\label{Scale_collapse}
\end{center}
\end{figure}

The solution in Eq. \ref{eq:a_of_t_general} represents a homogeneous evolution of the spherically symmetric distribution of the scalar field in Eq. \ref{eq:phi_of_t_general}. We plot $a(t)$ as a function of $t$ (taking $t_{0} = 0$) in Fig. \ref{Scale_collapse}. The graph on top is for different values of $p$, while $\eta$ and $\lambda$ are kept fixed. The graph in the middle is for for different values of $\eta$, while $p$ and $\lambda$ are kept fixed. The bottom graph is for different values of $\lambda$. It is interesting to note that the areal radius $a(t)$ never really goes to zero at any finite value of $t$. Depending on the parameter space, it is either an asymptotic collapse or an expanding solution with different rates of expansion at different time. For all collapsing solution plotted in Fig. \ref{Scale_collapse}, there is always an initial phase of expansion before the collapse.  \\ 

\begin{figure}[t!]
\begin{center}
\includegraphics[angle=0, width=0.40\textwidth]{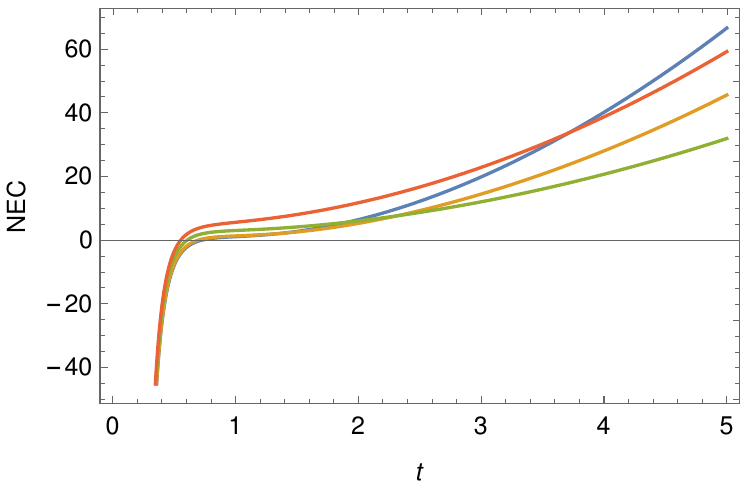}
\includegraphics[angle=0, width=0.40\textwidth]{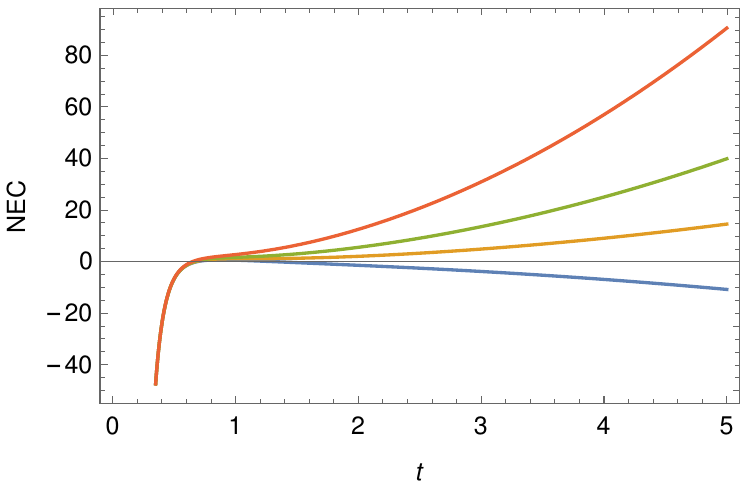}
\includegraphics[angle=0, width=0.40\textwidth]{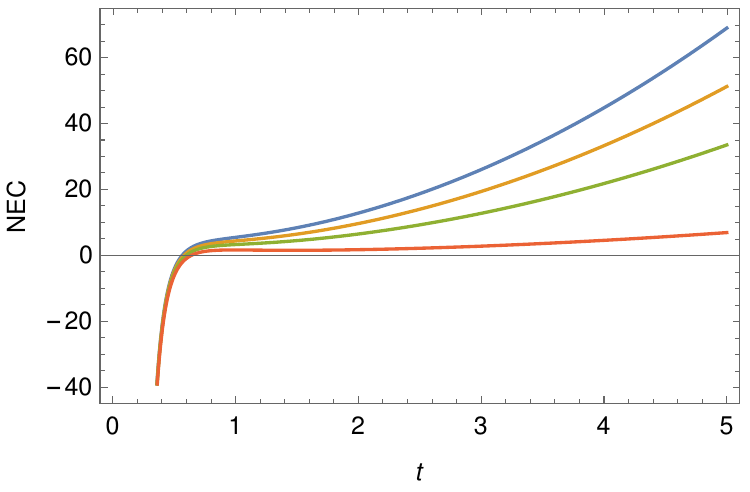}
\caption{Evolution of Null Energy Condition for the constituent fluid as a function of $t$. The graph on top is for different values of $p$, while $\eta$ and $\lambda$ are kept fixed. The graph in the middle is for for different values of $\eta$, while $p$ and $\lambda$ are kept fixed. The bottom graph is for different values of $\lambda$}
\label{NEC}
\end{center}
\end{figure}

\begin{figure}[t!]
\begin{center}
\includegraphics[angle=0, width=0.40\textwidth]{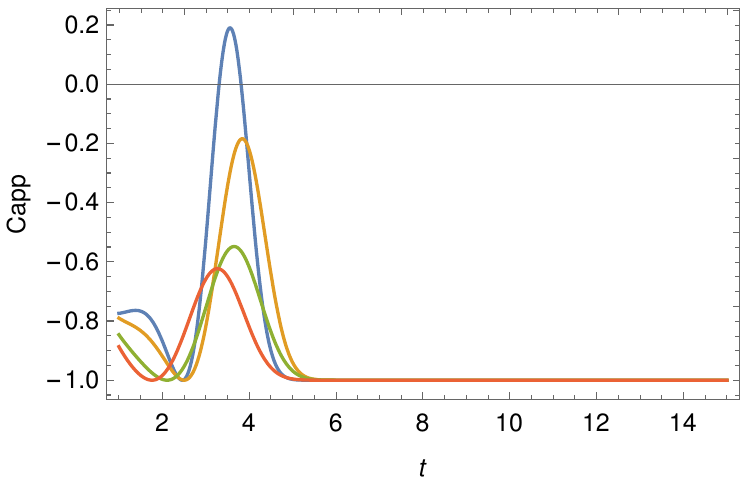}
\includegraphics[angle=0, width=0.40\textwidth]{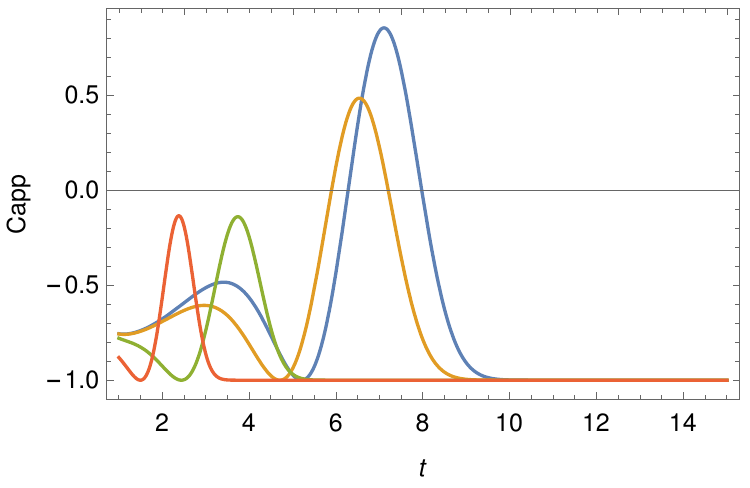}
\includegraphics[angle=0, width=0.40\textwidth]{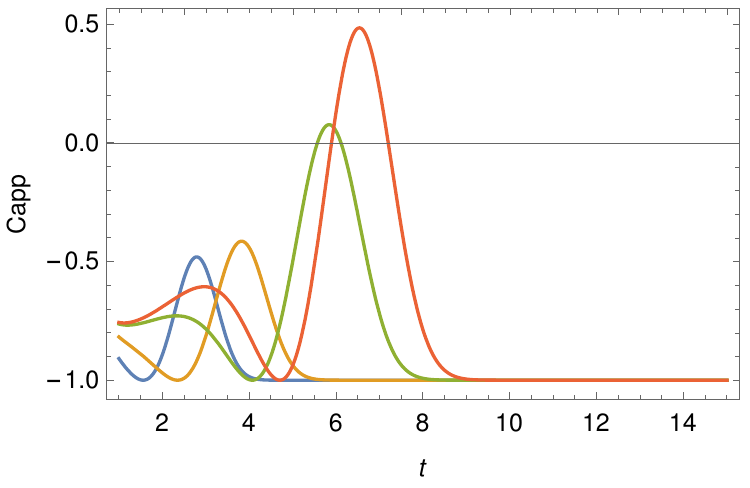}
\caption{The Apparent Horizon Condition as a function of $t$.  The graph on top is for different values of $p$, while $\eta$ and $\lambda$ are kept fixed. The graph in the middle is for for different values of $\eta$, while $p$ and $\lambda$ are kept fixed. The bottom graph is for different values of $\lambda$}
\label{Capp}
\end{center}
\end{figure}

\begin{figure}[t!]
\begin{center}
\includegraphics[angle=0, width=0.40\textwidth]{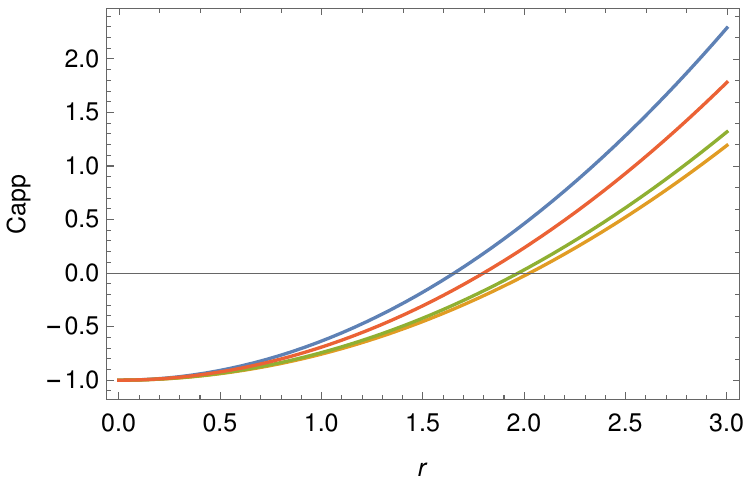}
\caption{The Apparent Horizon Condition as a function of $r$.}
\label{Cappr}
\end{center}
\end{figure}

The asymptotic nature of the collapse indicates that the constituent elements of the collapsing sphere are capable of generating an effective force opposing gravity. The existence of this effective force is understood from the energy conditions being obeyed/violated by the energy-momentum distribution. Our advantage is that we have chosen a solution out of the Chiellini integrability framework, applicable on the Klein-Gordon equation for the scalar field evolution alone. The rest of the field equations, namely, Eqs. (\ref{fe1minimal}) and (\ref{fe2minimal}) can therefore be used to derive the energy density and pressure of the fluid. In GR, the Null Energy Condition (NEC) is one of the fundamental
energy conditions imposed on the stress--energy tensor $T_{\mu\nu}$ to ensure physically reasonable matter distributions. It requires that for any null vector $k^\mu$ satisfying $k^\mu k_\mu = 0$, the energy--momentum tensor obeys $T_{\mu\nu} k^\mu k^\nu \geq 0$. For a perfect fluid with energy density $\rho$ and isotropic pressure $p$ the NEC gives
\begin{equation}
T_{\mu\nu} k^\mu k^\nu = (\rho + p)(u_\mu k^\mu)^2 \geq 0.
\end{equation}
Since $(u_\mu k^\mu)^2 > 0$ for any null vector $k^\mu$, the condition simplifies into
\begin{equation}
\rho + p \geq 0.
\end{equation}
Violation of the NEC often signals the presence of exotic matter or non-standard fields and it is closely related to the possibility
of phenomena such as cosmological bounces, wormholes or accelerated expansion in modified gravity theories. For the present case, using the energy density and pressure we calculate the NEC and plot it as a function of time in Fig. \ref{NEC}. It is quite clear from the graphs that the expanding solutions violate NEC and indicate a dark-energy like behavior. However, after the onset of collapse, the energy condition is well-satisfied. We also note that the scalar field remains canonical and satisfies all energy conditions. Using the exact solution for the scalar field, it is straightforward to check that
\begin{equation}
\phi(x)=\sqrt{x^{-2}+\sqrt{p/3}}, \qquad x=t-t_0,
\end{equation}
the scalar kinetic term behaves as
\begin{equation}
\dot{\phi}^2 = \frac{1}{x^4\left(1+x^2\sqrt{p/3}\right)},
\end{equation}
which is strictly positive for all values of time. Moreover, the kinetic term decays monotonically as $t \to \infty$. Therefore, it is straightforward to deduce from the scalar-field equation of state that the field can never lead to an effective negative pressure. \\

For a homogeneous and spatially flat collapsing spacetime described by $ds^2 = -dt^2 + a^2(t)\left(dr^2 + r^2 d\Omega^2\right)$, we can define the areal (or physical) radius as $R(t,r) = a(t)r$. We investigate the condition for a formation of apparent horizon, i.e., the requirement that the expansion of outgoing/ingoing null geodesics vanishes:
\begin{eqnarray}\nonumber
C_{app} \equiv g^{\mu\nu} \partial_\mu R \, \partial_\nu R 
&=& g^{tt}(\partial_t R)^2 + g^{rr}(\partial_r R)^2, \\
&=& -(\dot{a}r)^2 + 1 = 0.
\end{eqnarray}

We rewrite the above equation as a condition on $C_{app}$ and plot the evolution as a function of time in Fig. \ref{Capp}. The graph on top is for different values of $p$, while $\eta$ and $\lambda$ are kept fixed. The graph in the middle is for for different values of $\eta$, while $p$ and $\lambda$ are kept fixed. The bottom graph is for different values of $\lambda$. It is remarkable to note that depending on the parameter space, there is either no apparent horizon or two apparent horizons. We also plot $C_{app}$ as a function of the radial distance from the centre, in Fig. \ref{Cappr}.  \\

\section{Smooth Matching at a Boundary Hypersurface}
The collapse of a spherical distribution of scalar field and perfect fluid is not consistent unless it is consistently matched with a suitable exterior. We choose the exterior to be described by a generalized Vaidya spacetime which can be viewed as a specific instance of a more general, spherically symmetric line element,
\begin{eqnarray}
ds^2 &=& -e^{2\psi(v,r)}\left(1-\frac{2m(v,r)}{r}\right)dv^2 + 2\epsilon e^{\psi(v,r)}dvdr \nonumber\\
&& + r^2(d\theta^2+\sin^2\theta\,d\phi^2), \quad (\epsilon = \pm 1).
\label{generalized-vaidya}
\end{eqnarray}
This represents a mixture of Type-I and Type-II matter sources. Type-I matter admits one timelike and three spacelike eigenvectors of the energy--momentum tensor, whereas Type-II matter possesses double-null eigenvectors. The function $m(v,r)$ serves as a generalized mass function, corresponding to the total gravitational energy within a sphere of radius $r$ \cite{barrabes, viqar, wang}.  

For $\epsilon = +1$, the null coordinate $v$ represents the advanced Eddington time, so $r$ decreases along ingoing null rays of constant $v$. Conversely, $\epsilon = -1$ corresponds to an outgoing null congruence. Setting $\psi(v,r)=0$ gives the familiar generalized Vaidya geometry. Adopting $\epsilon=+1$, the metric reduces to
\begin{equation}
ds^2 = -\left(1-\frac{2m(v,r)}{r}\right)dv^2 + 2dvdr + r^2d\Omega^2.
\label{line-element}
\end{equation}

Choosing the pair of null vectors
\begin{eqnarray}\label{normalization}
&&l_{\mu} = \delta^0_{\mu}, \quad n_{\mu} = \frac{1}{2}\left(1-\frac{2m(v,r)}{r}\right)\delta^0_{\mu} - \delta^1_{\mu}, \\ &&
l_{\mu}l^{\mu}=n_{\mu}n^{\mu}=0, \quad l_{\mu}n^{\mu}=-1,
\end{eqnarray}
the corresponding energy--momentum tensor can be expressed as \cite{viqar, wang}
\begin{eqnarray}\label{EMT1}
&&T_{\mu\nu}=T^{(n)}_{\mu\nu}+T^{(m)}_{\mu\nu},\\ &&
T^{(n)}_{\mu\nu} = \mu l_{\mu}l_{\nu}, \quad
T^{(m)}_{\mu\nu} = (\rho+p)(l_{\mu}n_{\nu}+l_{\nu}n_{\mu})+pg_{\mu\nu}.
\label{EMT2}
\end{eqnarray}
The Einstein field equations then yield
\begin{eqnarray}\label{fieldeq}
\mu = \frac{2}{r^2}\frac{\partial m}{\partial v}, \quad
\rho = \frac{2}{r^2}\frac{\partial m}{\partial r}, \quad
p = -\frac{1}{r}\frac{\partial^2 m}{\partial r^2}.
\end{eqnarray}
Here, $T^{(n)}_{\mu\nu}$ represents radiation propagating along null hypersurfaces ($v=\text{const}$), whereas $T^{(m)}_{\mu\nu}$ corresponds to timelike fluid components. When $\rho = p = 0$, the metric reduces to the standard Vaidya form \cite{vaidya, vaidya1, vaidya2, vaidya3, vaidya4}.  \\

We now calculate the first and second fundamental forms \cite{deru, seno, santosmatching, chanmatching, sharp} and match a homogeneous FRW interior to a generalized Vaidya exterior across a boundary hypersurface $\Sigma$ defined by $r=r_0$. The interior and exterior line elements are
\begin{eqnarray}
ds_-^2 &=& -dt^2 + a^2(t)\left[dr^2 + r^2(d\theta^2+\sin^2\theta\,d\varphi^2)\right],
\label{inside metric}
\end{eqnarray}
and
\begin{eqnarray}
ds_+^2 &=& -\left(1-\frac{2M(r_v,v)}{r_v}\right)dv^2 + 2dvdr_v + r_v^2d\Omega^2,
\label{outside metric}
\end{eqnarray}
where $M(r_v,v)$ denotes the generalized mass function. The same hypersurface $\Sigma$ can be described in exterior coordinates by $r_v=R(t)$ and $v=T(t)$. Then, the induced metrics on $\Sigma$ from both sides become
\begin{eqnarray}
ds_{-,\Sigma}^2 &=& -dt^2 + a^2(t)r_0^2 d\Omega^2, \nonumber\\
ds_{+,\Sigma}^2 &=& -\left[\left(1-\frac{2M_{\Sigma}(t)}{R(t)}\right)\dot{T}^2 - 2\dot{T}\dot{R}\right]dt^2 + R^2(t)d\Omega^2, \nonumber
\end{eqnarray}
where $M_{\Sigma}(t)=M(R(t),T(t))$.  
Matching the first fundamental form ($ds_{-,\Sigma}^2 = ds_{+,\Sigma}^2$) gives
\begin{eqnarray}
&&\frac{dT}{dt} = \frac{1}{\sqrt{1-\frac{2M_{\Sigma}(t)}{R(t)} - 2\frac{dR}{dT}}}, 
\label{con 1}\\
&&R(t)=r_0a(t).
\label{con 2}
\end{eqnarray}

If the interior evolves as a homogeneous collapse, a reasonable assumption for the scale factor is $a(t)\sim a_0(t_0-t)^p$. To match the second fundamental form, we compute the normals to $\Sigma$ in both regions:
\begin{eqnarray}
&& n_-^t=0, \quad n_-^r=a(t), \quad n_-^{\theta}=n_-^{\varphi}=0, \\&& \label{inside normal}
n_+^v = \frac{1}{\sqrt{1-\frac{2M_{\Sigma}(t)}{R(t)} - 2\frac{dR}{dT}}}, \nonumber\\&&
n_+^{r_v} = \frac{1 - \frac{2M_{\Sigma}(t)}{R(t)} - \frac{dR}{dT}}
{\sqrt{1-\frac{2M_{\Sigma}(t)}{R(t)} - 2\frac{dR}{dT}}}, \quad
n_+^{\theta}=n_+^{\varphi}=0.
\label{outside normal}
\end{eqnarray}

From these, the non-zero components of the extrinsic curvature from the interior are
\begin{eqnarray}
K_{tt}^- = 0, \quad
K_{\theta\theta}^- = r_0a(t), \quad
K_{\varphi\varphi}^- = r_0a(t)\sin^2\theta,
\label{inside extrinsic}
\end{eqnarray}
and from the exterior,
\begin{eqnarray}
K_{tt}^+ &=& \frac{\partial M_{\Sigma}}{\partial R} - \frac{M_{\Sigma}}{r_0a(t)} - r_0^2a(t)\ddot{a}(t), \nonumber\\
K_{\theta\theta}^+ &=& R(t)\frac{1 - \frac{2M_{\Sigma}(t)}{R(t)} - \frac{dR}{dT}}
{\sqrt{1-\frac{2M_{\Sigma}(t)}{R(t)} - 2\frac{dR}{dT}}}, \nonumber\\
K_{\varphi\varphi}^+ &=& R(t)\sin^2\theta \frac{1 - \frac{2M_{\Sigma}(t)}{R(t)} - \frac{dR}{dT}}
{\sqrt{1-\frac{2M_{\Sigma}(t)}{R(t)} - 2\frac{dR}{dT}}}.
\label{outside extrinsic}
\end{eqnarray}

Equating the extrinsic curvatures from both sides leads to
\begin{eqnarray}
r_0a(t) = R(t)\frac{1 - \frac{2M_{\Sigma}(t)}{R(t)} - \frac{dR}{dT}}
{\sqrt{1-\frac{2M_{\Sigma}(t)}{R(t)} - 2\frac{dR}{dT}}},
\label{con 3}
\end{eqnarray}
and
\begin{eqnarray}
\frac{\partial M_{\Sigma}(t)}{\partial R(t)} = \frac{M_{\Sigma}(t)}{r_0a(t)} + r_0^2a(t)\ddot{a}(t).
\label{con 4}
\end{eqnarray}

Technically, Eqs.~(\ref{con 1})-(\ref{con 3}) define a generic set of identities required for a smooth matching at the boundary hypersurface. The particular conditions for our metric can easily be determined through the mass function at the boundary $\Sigma$.

\section{Conclusion}

In this work we have shown that the Chiellini integrability condition, originally developed in the study of nonlinear dissipative systems, can have natural applications in General Relativity through scalar fields. It is quite possible to reformulate the Klein-Gordon equation into such an integrable mechanism. We explore this possibility and construct exact solutions for a homogeneous and isotropic scalar field collapse. The derived exact solutions reveal that the areal radius monotonically decrease as a function of time but there is no formation of a zero proper-volume singularity at any finite time. The collapse is therefore asymptotic in nature. This behaviour originates from the interplay between the non-linear self-interaction potential and the effective damping term arising from the gravitational coupling. Unlike generic scalar-field collapse scenarios that culminate in finite-time singularities, the present integrable structure enforces a non-singular collapse. The associated energy-momentum tensor allows a direct examination of the energy conditions for the constituent elements. We check, in particular, that the perfect fluid present alongwith the scalar field can violate the Null energy condition during the collapse, although the scalar field remains canonical for all coordinate values. Furthermore, we analyze the trapped surface formation condition and find out that any formation of a trapped surface/apparent horizon depends sensitively on the parameter space : the spacetime may admit no apparent horizon or multiple horizons, depending on the choice of $(\lambda,\eta,p,c_1)$. This parameter-dependent horizon structure highlights the dynamical richness of the model while preserving an analytical control.  \\

We also carry out a consistent matching of the homogeneous interior with a generalized Vaidya exterior using the standard junction conditions. The resulting boundary conditions determine the effective mass function and ensure a smooth embedding of the collapsing configuration into an exterior radiating spacetime. This demonstrates that the integrable interior solution is not merely a formal construction but can be incorporated into a globally consistent gravitational-collapse scenario. Overall, the Chiellini integrability condition provides a mathematically robust framework for constructing exact scalar-field solutions in General Relativity. It establishes a concrete bridge between nonlinear integrable systems and relativistic gravitational dynamics, yielding an analytically controlled model of nonsingular collapse. A future work with a similar set of motivation can include extensions into anisotropic geometries, inhomogeneous scalar configurations or non-minimally coupled scalar field collapse scenarios, with the aim of further clarifying the role of integrable structures in gravitational physics.  \\

\paragraph*{Acknowledgement}
The authors acknowledge Vellore Institute of Technology for the financial support through its Seed Grant (No. SG20230027), 2023.

\end{document}